\documentclass[a4paper,12pt]{article}

\usepackage[left=2.5cm,right=2.5cm,top=2.5cm,bottom=2.5cm]{geometry}
\usepackage{graphicx}
\usepackage{amssymb}
\usepackage{multirow}
\usepackage{amsmath}
\usepackage{psfrag}
\usepackage{setspace}
\usepackage{subcaption}
\usepackage[colorlinks=true,bookmarks=true]{hyperref}
\usepackage{float}
\usepackage{cite}
\usepackage[utf8]{inputenc}

\hypersetup{citecolor=blue}
\newcommand{\dif}{\mathrm{d}}
\newcommand{\Eqref}[1]{(\ref{#1})}
\newcommand{\half}{\frac{1}{2}}
\newcommand{\expo}[1]{\mathrm{e}^{#1}}
\newcommand{\brac}[1]{\left(#1 \right)}
\newcommand{\sbrac}[1]{\left[#1\right]}

\newcommand{\barnab}{\bar{\nabla}}

\begin{document}

\title{Cohomogeneity-1 solutions in Einstein-Maxwell-dilaton gravity}
\author{Yen-Kheng Lim\footnote{E-mail: phylyk@nus.edu.sg}\\\textit{\normalsize{Department of Physics, National University of Singapore,}}\\\textit{\normalsize{Singapore 117551}}}
\date{\normalsize{\today}}
\maketitle

\begin{abstract}
  The field equations for Einstein-Maxwell-dilaton gravity in $D$ dimensions are reduced to an effective one-dimensional system under the influence of exponential potentials. Various cases where exact solutions can be found are explored. With this procedure, we present interesting solutions such as a one-parameter generalisation of the dilaton-Melvin spacetime and a three-parameter solution that interpolates between the Reissner-Nordstr\"{o}m and Bertotti-Robinson solutions. This procedure also allows simple, alternative derivations of known solutions such as the Lifshitz spacetime and the planar Anti-de Sitter naked singularity. In the latter case, the metric is cast in a simpler form which reveals the presence of an additional curvature singularity.
\end{abstract}

\section{Introduction} \label{intro}

The discovery of black rings, along with recent developments in string theory and various applications of the AdS/CFT correspondence has generated great interest in higher-dimensional spacetimes. Much of these situations of interest arise as solutions to Einstein-Maxwell-dilaton gravity with various scalar potentials. Indeed, the inclusion of Maxwell and dilaton fields complicates the equations of motion, where in higher dimensions they already lack the various solution-generating techniques available in the four-dimensional case. 

In four dimensions, despite the high non-linearity of the Einstein equations, there are a rich variety of methods developed to find static and stationary solutions. These solution-generating techniques are built upon the assumption of certain symmetries or the presence of Killing vectors in the metric.

In spacetimes with two commuting Killing vectors, the metric can be cast into the Weyl-Lewis-Papapetrou form. Ernst made use of this form to reformulate the Einstein equations in terms of complex potentials \cite{Ernst:1967wx}. This was subsequently generalised to include electromagnetic fields in Ref.~\cite{Ernst:1967by}. In this form, one can perform symmetry transformations to generate new solutions from known ones. For instance, one can embed a black hole in a background magnetic or electric field by performing a Harrison transformation \cite{Harrison:1968}. The simplest example of this is perhaps the magnetised Schwarzschild black hole found by Ernst \cite{Ernst:1976}.

Geroch \cite{Geroch:1970nt} has developed a more general analysis where the spacetime is required to carry only one Killing vector. This was recently extended to include a non-zero cosmological constant by Leigh et al.~\cite{Leigh:2014dja}, where they traced the symmetries broken by the presence of the cosmological constant and recast the equations of motion as a one-dimensional system which was solved using the Hamilton-Jacobi method. A further generalisation of this method to include electromagnetic fields was done by Klemm et al.~\cite{Klemm:2015uba}.

In higher dimensional vacuum gravity, $D$-dimensional static spacetimes with $(D-2)$ Killing vectors were characterised in terms of \emph{rod structures} by Emparan and Reall \cite{Emparan:2001wk}. This was generalised to stationary spacetimes by Harmark \cite{Harmark:2004rm,Harmark:2005vn}, and spacetimes with $U(1)\times U(1)$ isometry by Chen and Teo \cite{Chen:2010zu}. While the rod structure formalism is not possible in the presence of a cosmological constant, Armas et al. \cite{Armas:2011ed} nevertheless developed the \emph{domain structure} description (which generalises the rod into higher dimensions). In higher-dimensional Einstein-Maxwell gravity, various magnetised black holes have been found \cite{Ortaggio:2004kr}.

Most of the works mentioned above were mainly interested in pure vacuum Einstein gravity, with some extensions to include a cosmological constant and Maxwell fields. In this paper, we will be interested in the case where the Maxwell field has an additional coupling to a scalar dilaton field. In the case of zero cosmological constant, some symmetry properties the Einstein-Maxwell-dilaton equations have been studied. For instance, the Harrison transformation is also possible here and various interesting solutions have been generated in this manner \cite{Dowker:1993bt,Galtsov:1998mhf,Yazadjiev:2005gs,Yazadjiev:2013hxa}. The equations of motion for rotating black holes with equal angular momenta in odd dimensions have been studied within Einstein-Maxwell-dilaton gravity by \cite{Blazquez-Salcedo:2013wka,Blazquez-Salcedo:2013yba}. When the black hole has equal angular momentum, the symmetry of the solution is enhanced and the spacetime can be represented by a cohomogeneity-1 metric \cite{Kunduri:2006qa} and the equations of motion can be reduced to an effective one-dimensional system.

In this paper, we take an approach inspired by Refs.~\cite{Leigh:2014dja,Klemm:2015uba}, and assume that our metric is cohomogeneity-1 to cast the Einstein-Maxwell-dilaton equations for static spacetimes into an effective one-dimensional system. In Refs.~\cite{Leigh:2014dja,Klemm:2015uba}, the metric was assumed to have one Killing vector that is hypersurface orthogonal to a product space of the form $\mathbb{R}^1\times\Sigma_2$. The reduction to a one-dimensional system is achieved by taking $\Sigma_2$ to be a 2-sphere $S^2$ and letting all field variables depend only on the coordinate on $\mathbb{R}^1$. The coordinate system was carefully scaled such that the cosmological constant $\Lambda$ stands alone in the effective Lagrangian without any field-dependent factors. This allows $\Lambda$ to be interpreted as a constant of motion, but at the cost of the `kinetic' terms having exponential prefactors. 

In the present paper, we allow our metric to have any $D\geq 4$ dimensions, and $\Sigma_{(D-2)}$ (generalising $\Sigma_2$ to higher dimensions) to be an Einstein space that can have zero, positive, or negative normalised curvature which we denote by $k=0$, $\pm 1$. We choose a different coordinate system where the exponential factors are removed from the kinetic terms instead. In this case, $\Lambda$ now acquires the exponential factor, though this appears on equal footing with a term involving $k$ and the Maxwell field which we will parametrise by $q$.

Therefore, the effective Lagrangian describes a dynamical system under the influence of various exponential potentials parametrised by $\Lambda$, $k$, and $q$. When one or more of these potentials are zero, the system becomes exactly solvable. While a full classification of $(\Lambda,k,q)$ and their solutions is beyond the scope of this paper, we will report the particular choices of potentials that yield new solutions, or known solutions in newer and simpler forms.

The rest of this paper is organised as follows. In Sec.~\ref{action}, we derive the Einstein-Maxwell-dilaton equations under our choice of the metric ansatz. Subsequently we demonstrate in Sec.~\ref{example} how to derive the Schwarzschild and Fisher/Janis-Newman-Winicour solutions using the ansatz and equations of motion. In Sec.~\ref{noLambda} we obtain solutions for Einstein-Maxwell-dilaton theory in the case of zero cosmological constant. Solutions for non-zero cosmological constant are presented in Sec.~\ref{Lambda}. Finally, a discussion and closing remarks are given in Sec.~\ref{conclusion}. Appendix \ref{reduction} shows how our particular metric ansatz is chosen and the effective Lagrangian is derived.

\section{Action and equations of motion} \label{action}

The Einstein-Maxwell-dilaton theory we are considering is described by the action
\begin{align}
 I&=\frac{1}{16\pi}\int\dif^Dx\sqrt{-g}\brac{R-2\Lambda-\expo{-2\alpha\varphi}F^2-\brac{\nabla\varphi}^2},
\end{align}
where $F=\dif A$ is the Maxwell 2-form arising form a gauge potential $A$, and $\varphi$ is the scalar dilaton coupled to the Maxwell via the coupling parameter $\alpha$. 

Varying the action with respect to the metric, gauge potential, and dilaton gives the Einstein-Maxwell-dilaton equations
\begin{subequations}
\begin{align}
 R_{\mu\nu}=\frac{2\Lambda}{D-2}g_{\mu\nu}&+2\expo{-2\alpha\varphi}F_{\mu\lambda}{F_\nu}^\lambda-\frac{1}{D-2}\expo{-2\alpha\varphi}F^2g_{\mu\nu}+\nabla_\mu\varphi\nabla_\nu\varphi,\\
 \nabla_\lambda\brac{\expo{-2\alpha\varphi}F^{\lambda\nu}}&=0,\\
 \nabla^2\varphi+\alpha\expo{-2\alpha\varphi}F^2&=0.
\end{align}
\end{subequations}
In component form, the Maxwell 2-form is given by $F_{\mu\nu}=\partial_\mu A_\nu-\partial_\nu A_\mu$ and we denote $F^2=F_{\mu\nu}F^{\mu\nu}$. We take our metric to have at least one Killing vector $\partial_\sigma$ which we allow to be either time-like or space-like. We further take the hypersurface orthogonal to this Killing vector to be a warped product $\mathbb{R}^1\times\Sigma_{D-2}$, where $\Sigma_{D-2}$ is a $(D-2)$-dimensional Einstein space(time) of constant curvature. As mentioned in Sec.~\ref{intro}, this is simply the static case of the ansatz considered in Refs.~\cite{Leigh:2014dja,Klemm:2015uba}, generalised to include dilaton fields and higher dimensions.

With these considerations, our metric may be written as
\begin{align}
 \dif s^2&=\epsilon\expo{2U}\dif\sigma^2+\expo{\frac{2\Omega-2U}{D-3}}\brac{\expo{2\Omega}\dif\lambda^2+\hat{h}_{ij}\dif x^i\dif x^j}, \label{metric_ansatz}
\end{align}
where $\lambda$ is the coordinate that parametrises $\mathbb{R}^1$ and $\hat{h}_{ij}$ is the metric on $\Sigma_{D-2}$. To ensure that our full metric carries Lorentzian signature, if $\partial_\sigma$ is time-like, we take $\epsilon=-1$ and $\hat{h}_{ij}$ to be Euclidean, and if $\partial_\sigma$ is space-like, we take $\epsilon=+1$ and $\hat{h}_{ij}$ is understood to be Lorentzian. For the solutions presented below, we adopt the following notation for the coordinate $\sigma$: 
\begin{align}
 \epsilon\dif\sigma^2=\left\{
  \begin{array}{cl}
   -\dif t^2, & \mbox{for time-like }\partial_\sigma,\\
   \dif\phi^2, & \mbox{for space-like }\partial_\sigma.
  \end{array}\right.
\end{align}

Turning to the matter fields, we write the gauge potential and dilaton field in the form
\begin{align}
 A&=\sqrt{\frac{D-2}{2(D-3)}}\chi\,\dif\sigma,\nonumber\\
  \varphi&=\sqrt{\frac{D-2}{D-3}}\psi,\quad \alpha=\sqrt{\frac{D-3}{D-2}}a. \label{field_ansatz}
\end{align}
Therefore, our Einstein-Maxwell-dilaton system is determined by the functions $U$, $\Omega$, $\chi$, and $\psi$. In the following, we will assume that these functions depend only on $\lambda$.

The Einstein-Maxwell-dilaton equations under this ansatz have particularly simple forms, which are
\begin{subequations} \label{EOMS}
\begin{align}
 \dot{\chi}&=\epsilon q\expo{2a\psi+2U},\\
 \ddot{\psi}&=-\epsilon q^2 a\expo{2a\psi+2U},\\
 \ddot{U}&=-\epsilon q^2\expo{2a\psi+2U}-\frac{2\Lambda}{D-2}\expo{\frac{2(D-2)\Omega-2U}{D-3}},\\
 \ddot{\Omega}&=k(D-3)^2\expo{2\Omega}-2\Lambda\expo{\frac{2(D-2)\Omega-2U}{D-3}},
\end{align}
\end{subequations}
with an additional equation that follows from the trace of the Einstein equation,
\begin{align}
 \dot{U}^2-\dot{\Omega}^2+\epsilon\expo{-2a\psi-2U}\dot{\chi}^2+\dot{\psi}^2+k(D-3)^2\expo{2\Omega}-\frac{2(D-3)}{D-2}\Lambda\expo{\frac{2(D-2)\Omega-2U}{D-3}}=0. \label{constraint}
\end{align}
The equations in Eq.~\Eqref{EOMS} can be regarded as the Euler-Lagrange equations that follow from a Lagrangian
\begin{align}
 \mathcal{L}&=\half\sbrac{\dot{U}^2-\dot{\Omega}^2+\epsilon\expo{-2a\psi-2U}\dot{\chi}^2+\dot{\psi}^2-k(D-3)^2\expo{2\Omega}+\frac{2(D-3)}{D-2}\Lambda\expo{\frac{2(D-2)\Omega-2U}{D-3}}},\label{Lagrangian}
\end{align}
and Eq.~\Eqref{constraint} can be regarded as a Hamiltonian constraint for this system. In Appendix \ref{reduction}, we show the line of reasoning that led us to the choice of the ansatz \Eqref{metric_ansatz} which results in the compact equations as they appear in \Eqref{EOMS}. 

Henceforth, our primary focus will be solving the system described by Eq.~\Eqref{Lagrangian} under the constraint \Eqref{constraint}. The metric, along with the Maxwell and dilaton fields are then reconstructed according to Eqs.~\Eqref{metric_ansatz} and \Eqref{field_ansatz}.

In summary, with the cohomogeneity-1 ansatz \Eqref{metric_ansatz}, the problem has been reduced to a one-dimensional Lagrangian with exponential potentials. We will show below that despite the simplicity of the equations of motion, this ansatz is sufficiently non-trivial where it is possible to construct various black-hole and electro-vacuum solutions.

Before proceeding to find solutions in the following sections, a few general comments are in order. We first note that $\chi$ is a cyclic variable of the Lagrangian such that $\dot{\chi}$ can be expressed in terms of $U$, $\psi$, and a constant of motion, $q$. Therefore the presence of $\chi$ behaves like an effective `interaction potential' between $U$ and $\Omega$. Similarly, $\Lambda$ can be regarded as parametrising the strength of the `interaction' between $U$ and $\Omega$, while $k$ parametrises a single exponential potential for $\Omega$. If the couplings can be removed either by setting the potentials to zero and/or by finding an appropriate change of variables, the equations of motion reduce to (copies of) Liouville's differential equation\footnote{A similar type of reduction to Liouville's equation has been used to find black holes in gravity with multiple Maxwell-dilaton fields \cite{Lu:2013uia}.}, which is exactly solvable. In the subsequent sections below we will explore various cases where this is possible.

In the case where any of the two variables are cyclic, simple symmetry transformations that preserve the Lagrangian allow us to generate new solutions from known ones. For example, in the absence of the cosmological constant and electromagnetic fields, the Lagrangian and constraint take the forms
\begin{align}
 \mathcal{L}&=\half\sbrac{\dot{U}^2-\dot{\Omega}^2+\dot{\psi}^2-k(D-3)^2\expo{2\Omega}},\nonumber\\
    0&=\dot{U}^2-\dot{\Omega}^2+\dot{\psi}^2+k(D-3)^2\expo{2\Omega}.
\end{align}
We see that this system is invariant under the transformation
\begin{align}
 U'&=U\cos\theta-\psi\sin\theta,\nonumber\\
 \psi'&=U\sin\theta+\psi\cos\theta, \nonumber\\
 \Omega'&=\Omega,\label{O(2)}
\end{align}
for some real parameter $\theta$. Therefore, with a known solution corresponding to $(U,\psi,\Omega)$, one can generate a new solution $(U',\psi',\Omega')$ with the transformation \Eqref{O(2)}.

\section{Example: The Schwarzschild and Fisher/JNW solutions} \label{example}

To demonstrate how solving Eqs.~\Eqref{Lagrangian} and \Eqref{constraint} gives a solution to the the field equations, we shall consider a simple case without a cosmological constant or a Maxwell field. This essentially leaves us with pure Einstein gravity sourced by a mass-less scalar field. In this case, let us take $k=1$ where $\Sigma_{D-2}$ is now a sphere $S^{D-2}$. We also take the Killing vector $\partial_\sigma$ to be time-like so that $\epsilon\dif\sigma^2=-\dif t^2$. The Lagrangian and constraint are
\begin{align}
 \mathcal{L}&=\half\sbrac{\dot{U}^2-\dot{\Omega}^2+\dot{\psi}^2-(D-3)^2\expo{2\Omega}},\label{Sch_Lagrangian}\\
 0&=\dot{U}^2-\dot{\Omega}^2+\dot{\psi}^2+(D-3)^2\expo{2\Omega}. \label{Sch_constraint}
\end{align}
The Euler-Lagrange equations are then
\begin{align}
 \ddot{\Omega}&=(D-3)^2\expo{2\Omega},\quad \dot{U}=\mathrm{constant}\equiv -b_1,\quad \dot{\psi}=\mathrm{constant}\equiv p,
\end{align}
which are solved by 
\begin{align}
 U&=-b_1\lambda+U_0,\nonumber\\
 \Omega&=-b_2(\lambda+\lambda_0)-\ln\brac{\frac{(D-3)^2}{4b_2^2c}-c\expo{-2b_2(\lambda+\lambda_0)}},\nonumber\\
 \psi&=p\lambda+\psi_0. \label{Sch_soln0}
\end{align}
We can simplify the expressions by shifting the zeros of $\lambda$, $\psi$, and $U$ such that $\frac{\expo{2b\lambda_0}}{c^2}=\frac{4b_2^2}{(D-3)^2}$ and $U_0=\psi_0=0$. 

We first seek the vacuum Schwarzschild solution by setting $p=0$, thereby switching off the scalar field. In this case, the constraint equation requires that $b_1=b_2\equiv b$. The solution is then
\begin{align}
 U&=-b_1\lambda,\quad \Omega=-\ln\brac{\frac{(D-3)\sinh b_2\lambda}{b_2}},\quad \psi=0. \label{Sch_soln}
\end{align}
The metric is reconstructed as
\begin{align}
 \dif s^2&=-\expo{-2b\lambda}\dif t^2+\brac{\frac{b\expo{b\lambda}}{(D-3)\sinh b\lambda}}^{\frac{2}{D-3}}\frac{b^2\;\dif\lambda^2}{(D-3)\sinh^2 b\lambda}\nonumber\\
         &\hspace{2cm}+\brac{\frac{b\expo{b\lambda}}{(D-3)\sinh b\lambda}}^{\frac{2}{D-3}}\dif\Omega^2_{(D-2)},
\end{align}
where we have denoted the metric on $\Sigma_{D-2}=S^{D-2}$ as $\hat{h}_{ij}\dif x^i\dif x^j=\dif\Omega_{(D-2)}^2$. To see that this is actually the $D$-dimensional Schwarzschild-Tangherlini metric, we perform the following transformation and redefinition of $b$:
\begin{align}
 \expo{-2b\lambda}=1-\frac{\mu}{r^{D-3}},\quad b=\half(D-3)\mu. \label{Sch_coordinates}
\end{align}
Then we find that the metric is now
\begin{align}
 \dif s^2&=-\brac{1-\frac{\mu}{r^{D-3}}}\dif t^2+\brac{1-\frac{\mu}{r^{D-3}}}^{-1}\dif r^2+r^2\dif\Omega^2_{(D-2)},
\end{align}
which is indeed the $D$-dimensional Schwarzschild-Tangherlini metric with mass parameter $\mu$.

Next, we note that the Lagrangian \Eqref{Sch_Lagrangian} and constraint \Eqref{Sch_constraint} are invariant under the $O(2)$ transformation \Eqref{O(2)}. Therefore, if we take $U$, $\Omega$, and $\psi$ given in Eq.~\Eqref{Sch_soln} as the seed, a new solution can be generated with Eq.~\Eqref{O(2)} to obtain
\begin{align}
 U'&=-\nu b\lambda,\quad\psi'=-\sqrt{1-\nu^2}b\lambda,\quad \Omega'=\Omega=-\ln\brac{\frac{(D-3)\sinh b_2\lambda}{b_2}},
\end{align}
where we have defined $\nu\equiv\cos\theta$. Reconstructing the metric and the scalar field, and further applying the transformation defined in Eq.~\Eqref{Sch_coordinates}, we have
\begin{align}
 \dif s^2&=-f^\nu\dif t^2+f^{\frac{1-\nu}{D-3}}\brac{\frac{\dif r^2}{f}+r^2\dif\Omega^2_{(D-2)}},\nonumber\\
   \varphi&=\half\sqrt{\frac{(D-2)(1-\nu^2)}{D-3}}\ln f,\quad f=1-\frac{\mu}{r^{D-3}},
\end{align}
which is the Fisher/Janis-Newman-Winicour (JNW)\footnote{In the literature, this solution is widely attributed to Janis, Newman, and Winicour \cite{Janis:1968zz}, though it turns out that it is a rediscovery of a solution obtained by Fisher \cite{Fisher:1948yn}.} solution in arbitrary dimensions \cite{Xanthopoulos:1989kb}. The generation of this spacetime via symmetry transformation from a Schwarzschild seed using Eq.~\Eqref{O(2)} was previously performed by Abdolrahimi and Shoom \cite{Abdolrahimi:2009dc}.

\section{Solutions with zero cosmological constant} \label{noLambda}

\subsection{Decoupling the equations of motion}

In this section, we shall consider Einstein-Maxwell-dilaton gravity without a cosmological constant. With $\Lambda=0$ in the Lagrangian \Eqref{Lagrangian}, $\Omega$ is decoupled from the other fields and may be solved exactly. However, if $a$ is non-zero, the presence of $\chi$ couples $U$ to $\psi$. Nevertheless we can decouple the equations of motion by introducing the transformation
\begin{align}
 U&=\frac{\xi-a\eta}{1+a^2},\quad\psi=\frac{\eta+a\xi}{1+a^2},\quad\chi=\frac{\zeta}{\sqrt{1+a^2}},
\end{align}
and the system is now
\begin{align}
 \mathcal{L}&=\half\sbrac{\frac{\dot{\xi}^2+\dot{\eta}^2+\epsilon\expo{-2\xi}\dot{\zeta}^2}{1+a^2}-\dot{\Omega}^2-k(D-3)^2\expo{2\Omega}},\label{EMd0_Lagrangian}\\
 0&=\frac{\dot{\xi}^2+\dot{\eta}^2+\epsilon\expo{-2\xi}\dot{\zeta}^2}{1+a^2}-\dot{\Omega}^2+k(D-3)^2\expo{2\Omega}.\label{EMd0_constraint}
\end{align}
with the Euler-Lagrange equations
\begin{subequations}\label{EMd0_EOM}
\begin{align}
 \dot{\zeta}&=\epsilon q\expo{2\xi},\\
 \dot{\eta}&=p,\\
 \ddot{\xi}&=-\epsilon q^2\expo{2\xi},\\
 \ddot{\Omega}&=k(D-3)^2\expo{2\Omega},
\end{align}
\end{subequations}
where $q$ and $p$ are constants that follow from the first integral of $\zeta$ and $\eta$.

\subsection{Reissner-Nordstr\"{o}m/Bertotti-Robinson interpolating solution}

Taking the case $\epsilon=-1$ and $k=1$, the solution to Eq.~\Eqref{EMd0_EOM} is
\begin{subequations}\label{RNBR_soln}
\begin{align}
 \xi&=-b_1\lambda-\ln\brac{\frac{q^2}{4cb_1^2}-c\expo{-2b_1\lambda}},\\
 \Omega&=-\ln\brac{\frac{(D-3)\sinh b_2\lambda}{b_2}},\\
 \zeta&=\frac{q}{2b_1c\brac{\frac{q^2}{4b_1^2c}-c\expo{-2b_1\lambda}}},\\
 \eta&=p\lambda.
\end{align}
\end{subequations}
where we have used a similar argument that took us from Eq.~\Eqref{Sch_soln0} to Eq.~\Eqref{Sch_soln} to cast $\Omega$ into the form above. Substituting Eq.~\Eqref{RNBR_soln} into the constraint \Eqref{EMd0_constraint} leads us to
\begin{align}
 b_1^2+p^2=(1+a^2)b_2^2,
\end{align}
thereby constraining one of the integration constants.

Reconstructing the solution, we have
\begin{align}
 \dif s^2&=-\frac{\expo{-\frac{2(b_1+ap)\lambda}{1+a^2}}\;\dif t^2}{\brac{\frac{q^2}{4b_1^2c}-c\expo{-2b_1\lambda}}^\frac{2}{1+a^2}}+\brac{\frac{q^2}{4b_1^2c}-c\expo{-2b_1\lambda}}^\frac{2}{(D-3)(1+a^2)}\sbrac{\frac{b_2\expo{\frac{(b_1+ap)\lambda}{1+a^2}}}{(D-3)\sinh b_2\lambda}}^\frac{2}{D-3}\nonumber\\
  &\quad\quad\times\brac{\frac{b_2^2\;\dif\lambda^2}{(D-3)^2\sinh^2b_2\lambda}+\dif\Omega^2_{(D-3)}},\nonumber\\
   A&=\sqrt{\frac{D-2}{2(D-3)\brac{1+a^2}}}\frac{q^2}{2b_1c\brac{\frac{q^2}{4b_1^2c}-c\expo{-2b_1\lambda}}}\;\dif t,\nonumber\\
   \varphi&=\sqrt{\frac{D-2}{D-3}}\frac{1}{1+a^2}\sbrac{(p-b_1a)\lambda-a\ln\brac{\frac{q^2}{4b_1^2c}-c\expo{-2b_1\lambda}}},\nonumber\\
   \alpha&=\sqrt{\frac{D-3}{D-2}}a,\quad b_1^2+p^2=(1+a^2)b_2^2.
\end{align}
We will show that this solution contains limits to the Reissner-Nordstr\"{o}m and Bertotti-Robinson solutions. The metric will appear much simpler if we consider the case $p=ab_1$, for which the constraint leads to $b_1=b_2\equiv b$, and this allows us to further apply the transformation \Eqref{Sch_coordinates}, which swaps $\lambda$ and $b$ to the more familiar, Schwarzschild-like conventions $r$ and $\mu$. The resulting solution is
\begin{align}
 \dif s^2&=-\frac{f}{H^{\frac{2}{1+a^2}}}\dif t^2+H^{\frac{2}{(D-3)(1+a^2)}}\brac{\frac{\dif r^2}{f}+r^2\dif\Omega^2_{(D-2)}},\nonumber\\
 A&=\sqrt{\frac{D-2}{2(D-3)\brac{1+a^2}}}\;\frac{q}{(D-3)\mu cH}\;\dif t,\quad\varphi=-\sqrt{\frac{D-2}{D-3}}\frac{a}{1+a^2}\ln H,\nonumber\\
 f&=1-\frac{\mu}{r^{D-3}},\quad H=\frac{q^2}{c\mu^2(D-3)^2}-c+\frac{c\mu}{r^{D-3}},\quad \alpha=\sqrt{\frac{D-3}{D-2}}a.\label{RNBR}
\end{align}
Fixing $p$ now leaves us with three independent parameters, $\mu$, $q$, and $c$.

We are now in a position to obtain the Reissner-Nordstr\"{o}m and Bertotti-Robinson limits from the above solution. First, if $c$ takes the value
\begin{align}
 c=-\half\pm\frac{\sqrt{\mu^2(D-3)^2+4q^2}}{2\mu(D-3)},
\end{align}
the function $H$ as defined in Eq.~\Eqref{RNBR} takes the form
\begin{align}
 H=1+\frac{c\mu}{r^{D-3}},
\end{align}
and the solution is simply the dilatonic Reissner-Nordstr\"{o}m black hole written in $p$-brane coordinates.

On the other hand, if $c$ takes the value
\begin{align}
 c=\frac{q}{\mu(D-3)}\equiv\frac{e}{\mu},\quad
\end{align}
and further transforming
\begin{align}
 t\rightarrow \frac{e^{\frac{2}{1+a^2}}}{D-3}t,\quad r\rightarrow\rho^{-\frac{1}{D-3}},
\end{align}
the solution becomes
\begin{align}
 \dif s^2&=-\brac{\frac{e}{\rho}}^{\frac{2}{1+a^2}}\frac{(1-\mu\rho)\dif t^2}{(D-3)^2}\nonumber\\
    &\quad+e^\frac{2}{(D-3)(1+a^2)}\rho^{-\frac{2a^2}{(D-3)(1+a^2)}}\brac{\frac{\dif\rho^2}{(D-3)^2\rho^2(1-\mu\rho)}+\dif\Omega^2_{(D-2)}},\nonumber\\
 A&=\sqrt{\frac{D-2}{2(D-3)}}\frac{e^{\frac{1-a^2}{1+a^2}}}{(D-3)\rho}\;\dif t,\quad\varphi=-\sqrt{\frac{D-2}{D-3}}\frac{a}{1+a^2}\ln e\rho.
\end{align}
We can recover the Bertotti-Robinson metric  in $D=4$ Einstein-Maxwell gravity by setting $a=0$,
\begin{align}
 \dif s^2&=\frac{e^2}{\rho^2}\brac{-(1-\mu\rho)\dif t^2+\frac{\dif\rho^2}{1-\mu\rho}+\rho^2\dif\Omega^2_{(D-2)}},\nonumber\\
    A&=\frac{e}{\rho}\dif t,
\end{align}
where the Poincar\'{e} slices in the $\mathrm{AdS}_2$ are written in Rindler-like coordinates. Similar interpolating solutions in $D=4$ Einstein-Maxwell gravity has been found by Halilsoy \cite{Halilsoy:1993}, and in $D=4$ Einstein-Maxwell-dilaton gravity with a Liouville potential by Mazharimousavi et al. \cite{Mazharimousavi:2009vh}.

\subsection{Generalised dilaton-Melvin spacetime}

We now consider the case where the Killing vector is space-like and $k=0$, so that $\hat{h}_{ij}$ is a $(D-2)$-dimensional Minkowski spacetime and we set $\epsilon\dif\sigma^2=\dif\phi^2$. A particular solution to Eq.~\Eqref{EMd0_EOM} in this case is
\begin{align}
 \xi&=-\lambda-\ln\brac{\frac{q^2}{4}+\expo{-2\lambda}},\quad\zeta=\frac{\half q\expo{2\lambda}}{1+\frac{q^2}{4}\expo{2\lambda}},\quad \gamma=\lambda,\quad\eta=0.\label{dilaton-Melvin_seed}
\end{align}
Before reconstructing the metric, we observe that for $k=0$, the Lagrangian \Eqref{EMd0_Lagrangian} and constraint \Eqref{EMd0_constraint} are invariant under the transformation
\begin{align}
 \eta'&=\eta\cosh\theta-\gamma\sinh\theta,\nonumber\\
 \gamma'&=-\eta\sinh\theta+\gamma\cosh\theta,
\end{align}
with $\xi$ and $\zeta$ remain unchanged. Taking Eq.~\Eqref{dilaton-Melvin_seed} as a seed, we now have\footnote{Of course, one could have also obtained this solution by choosing a nonzero solution for $\eta$ in \Eqref{dilaton-Melvin_seed} and choosing the appropriate constants.}
\begin{align}
 \xi&=-\lambda-\ln\brac{\frac{q^2}{4}+\expo{-2\lambda}},\quad\zeta=\frac{q\expo{2\lambda}}{2\brac{1+\frac{q^2}{4}\expo{2\lambda}}},\quad \gamma'=\nu\lambda,\quad\eta'=-\sqrt{\nu^2-1}\lambda,
\end{align}
where we have introduced the parametrisation $\nu=\cosh\theta$.

Reconstructing the solution, with the transformation $\rho=\expo{\lambda}$, we have a one-parameter generalisation of the dilaton-Melvin spacetime:
\begin{align}
 \dif s^2&=\rho^{\frac{2\brac{1+a\sqrt{\nu^2-1}}}{1+a^2}}H^{-\frac{2}{1+a^2}}\dif\phi^2\nonumber\\
    &\quad+\rho^{\frac{2\brac{\nu\sqrt{1+a^2}-1-a\sqrt{\nu^2-1}}}{(1+a^2)(D-3)}}H^{\frac{2}{(1+a^2)(D-3)}}\brac{\rho^{\frac{2\nu-2\sqrt{1+a^2}}{\sqrt{1+a^2}}}\dif\rho^2-\dif t^2+\dif\vec{x}^2_{(D-3)}},\nonumber\\
   A&=\sqrt{\frac{D-2}{2(D-3)\brac{1+a^2}}}\;\frac{q\rho^2}{2H}\;\dif\phi,\quad\expo{-2\alpha\varphi}=\rho^{\frac{2a\sqrt{\nu^2-1}-2a^2}{1+a^2}}H^{\frac{2a^2}{1+a^2}},\nonumber\\
   H&=1+\frac{1}{4}q^2\rho^2,\quad\alpha=\sqrt{\frac{D-3}{D-2}}a,
\end{align}
where we have denoted $\hat{h}_{ij}\dif x^i\dif x^j=-\dif t^2+\dif\vec{x}_{(D-3)}^2$. For the special case of $\nu=\sqrt{1+a^2}$, the solution reduces to
\begin{align}
 \dif s^2&=\rho^2H^{-\frac{1}{1+a^2}}\dif\phi^2+H^{\frac{2}{(D-3)(1+a^2)}}\brac{\dif\rho^2-\dif t^2+\dif\vec{x}^2_{(D-3)}},\nonumber\\
 A&=\sqrt{\frac{D-2}{2(D-3)\brac{1+a^2}}}\;\frac{q\rho^2}{2H}\;\dif\phi,\quad\expo{-2a\psi}=H^{\frac{2a^2}{1+a^2}},
\end{align}
where $H$ and $a$ are the same as above. This is the familiar dilaton-Melvin solution considered in Refs.~\cite{Gibbons:1987ps,Yazadjiev:2005gs}.

\section{Solutions with a non-zero cosmological constant} \label{Lambda}

\subsection{Planar AdS naked singularity}

In this section, we turn to the case with a non-zero cosmological constant and zero Maxwell fields. Therefore the solutions in this case correspond to (Anti-)de Sitter-scalar theory. Here we shall consider the case $k=0$, and a time-like $\partial_\sigma$. Therefore we use the notation $\hat{h}_{ij}\dif x^i\dif x^j=\dif\vec{x}_{(D-2)}^2$ and $\epsilon\dif\sigma^2=-\dif t^2$. The Lagrangian and constraint equations are
\begin{align}
 \mathcal{L}&=\half\sbrac{\dot{U}^2-\dot{\Omega}^2+\dot{\psi}^2+\frac{2(D-3)}{D-2}\Lambda\expo{\frac{2(D-2)\Omega-2U}{D-3}}}, \label{PlanarAdS_Lagrangian}\\
  0&=\dot{U}^2-\dot{\Omega}^2+\dot{\psi}^2-\frac{2(D-3)}{D-2}\Lambda\expo{\frac{2(D-2)\Omega-2U}{D-3}}.\label{PlanarAdS_constraint}
\end{align}
We see that the presence of the cosmological constant couples $\Omega$ to $U$. Nevertheless, if we introduce the transformation
\begin{align}
 U&=\frac{\beta-(D-2)\alpha}{D-1},\quad\Omega=\frac{(D-2)\beta-\alpha}{D-1},\quad\psi=\sqrt{\frac{D-3}{D-1}}\gamma,
\end{align}
the Lagrangian and constraint become
\begin{align}
 \mathcal{L}&=\frac{D-3}{2(D-1)}\brac{\dot{\alpha}^2-\dot{\beta}^2+\dot{\gamma}^2+K\expo{2\beta}},\label{PlanarAdS_Lagrangian2}\\
 0&=\dot{\alpha}^2-\dot{\beta}^2+\dot{\gamma}^2-K\expo{2\beta},\label{PlanarAdS_constraint2}
\end{align}
where $K=2(D-1)\Lambda/(D-2)$. 

This system now appears in the same form as Eqs.~\Eqref{Sch_Lagrangian} and \Eqref{Sch_constraint} with a slightly different potential strength $K$. It also has a similar invariance under the $O(2)$ transformation \Eqref{O(2)}. We can therefore carry over the results for the Fisher/JNW solution in Sec.~\ref{example} to our present case, giving us
\begin{align}
 \alpha=\nu c\lambda,\quad\beta=-\ln\brac{\frac{\sqrt{K}\sinh c\lambda}{c}},\quad\gamma=\sqrt{1-\nu^2}c\lambda,\quad |\nu|\leq 1.
\end{align}
Introducing a similar transformation to the $r$ coordinates defined in \Eqref{Sch_coordinates}, and further setting $2\Lambda=-(D-1)(D-2)\ell^{-2}$, the full solution is reconstructed upon an appropriate rescaling of $t$ and $\mu$ to give
\begin{align}
 \dif s^2&=-\frac{r^2}{\ell^2}f^{\frac{\nu(D-2)+1}{D-1}}\dif t^2+\frac{\ell^2}{r^2}\frac{\dif r^2}{f}+\frac{r^2}{\ell^2}f^{\frac{1-\nu}{D-1}}\dif\vec{x}^2_{(D-2)},\nonumber\\
 \varphi&=\half\sqrt{\frac{(D-2)\brac{1-\nu^2}}{D-1}}\ln f,\quad f=1-\frac{\mu}{r^{D-1}}. \label{naked_AdS}
\end{align}
This is the metric found recently by Saenz and Martinez \cite{Saenz:2012ga}. To see this, we introduce the transformation
\begin{align}
 r^{D-1}=x+b,\quad b=\frac{\mu}{2}. \label{saenz_coordinates}
\end{align}
The solution then becomes
\begin{align}
 \dif s^2&=-\ell^{-2}(x+b)^{\frac{1-\nu(D-2)}{D-1}}(x-b)^{\frac{1+\nu(D-2)}{D-1}}\dif t^2+\frac{\ell^2\dif x^2}{(D-1)^2(x^2-b^2)}\nonumber\\
   &\quad+\ell^2\brac{x+b}^{\frac{1+\nu}{D-1}}\brac{x-b}^{\frac{1-\nu}{D-1}}\dif\vec{x}^2_{(D-2)},\nonumber\\
   \varphi&=\half\sqrt{\frac{(D-2)(1-\nu^2)}{D-1}}\ln\frac{x-b}{x+b},
\end{align}
which is precisely the form given in Eqs.~(13) and (14) in Ref.~\cite{Saenz:2012ga}. In their paper, Saenz and Martinez \cite{Saenz:2012ga} pointed out curvature singularity at $x=b$, which by Eq.~\Eqref{saenz_coordinates}, corresponds to $r=\mu$ in the form given in Eq.~\Eqref{naked_AdS}. Additionally, from Eq.~\Eqref{naked_AdS}, we see that there is yet another curvature singularity located at $r=0$.\footnote{This was possibly ignored in \cite{Saenz:2012ga} because it was located beyond their coordinate range of interest.}

From the form in $r$ coordinates given in Eq.~\Eqref{naked_AdS}, we easily obtain the asymptotic Anti-de Sitter (AdS) limit for large $r$. Within the same coordinates we can set $\nu=1$ to get the planar AdS black hole, where the $r=\mu$ surface is simply a black hole horizon concealing the yet remaining curvature singularity at $r=0$. In light of this, we can interpret this solution as the planar AdS analogue to the Fisher/JNW solution, where a Schwarzschild black hole dressed with a scalar field turns its horizon into a curvature singularity. Similarly, we see that dressing a planar AdS black hole with a scalar field turns its $r=\mu$ horizon into a curvature singularity.

\subsection{Lifshitz spacetime}

In this section, we shall use the Lagrangian \Eqref{Lagrangian} and constraint \Eqref{constraint} to provide a simple derivation of the Lifshitz spacetime \cite{Kachru:2008yh,Taylor:2008tg,Pang:2009ad,Park:2013goa}. This spacetime is of interest in the context of gauge/gravity duality where its holographic dual is a non-relativistic field theory \cite{Taylor:2015glc}. While this spacetime is already well known, it is hoped that the simplicity of Eq.~\Eqref{Lifshitz_EOM} below provides a useful tool for further studies of condensed matter systems and their gravity duals, particularly those that require Lifshitz asymptotics.

As in the previous section, we consider the case $k=0$ and $\epsilon\dif\sigma^2=-\dif t^2$. The equations of motion in this case are 
\begin{subequations} \label{Lifshitz_EOM}
\begin{align}
 \dot{\chi}&=-q\expo{2a\psi+2U},\\
 \ddot{U}&=q^2\expo{2a\psi+2U}-\frac{2\Lambda}{D-2}\expo{\frac{2(D-2)\Omega-2U}{D-3}},\\
 \ddot{\Omega}&=-2\Lambda\expo{\frac{2(D-2)\Omega-2U}{D-3}},\\
 \ddot{\psi}&=q^2a\expo{2a\psi+2U},
\end{align}
\end{subequations}
along with the constraint
\begin{align}
 \dot{U}^2-\dot{\Omega}^2+q^2\expo{2a\psi+2U}+\dot{\psi}^2-\frac{2(D-3)}{D-2}\Lambda\expo{\frac{2(D-2)\Omega-2U}{D-3}}=0.
\end{align}
While the equations of motion are coupled, we can find a non-trivial solution by assuming that $\Omega$ and $\psi$ are proportional to $U$. In anticipation of the Lifshitz solution, we choose a parametrisation of the proportionality constants such that
\begin{align}
 \Omega=\frac{D-3+\nu}{\nu}U,\quad\psi=\frac{D-2}{a\nu}U,
\end{align}
for $\nu\geq 1$. The equations of motion now become
\begin{align}
 \ddot{U}&=\frac{\nu a^2q^2}{D-2}\expo{\frac{2(D-2+\nu)U}{\nu}}=-\frac{2\Lambda\nu}{D-3+\nu}\expo{\frac{2(D-2+\nu)U}{\nu}}=\brac{q^2-\frac{2\Lambda}{D-2}}\expo{\frac{2(D-2+\nu)U}{\nu}}.
\end{align}
Consistency then requires $q$ and $a$ to satisfy
\begin{align}
 q^2&=-\frac{2\Lambda(\nu-1)(D-3)}{(D-2)(D-3+\nu)},\quad a^2=\frac{(D-2)^2}{(D-3)(\nu-1)}.
\end{align}
Anticipating the Lifshitz solution again, we further parametrise the cosmological constant by $-2\Lambda=(D-2+\nu)(D-3+\nu)\ell^{-2}$. A solution for $U$ that satisfies the constraint is
\begin{align}
 U&=-\frac{\nu}{D-2+\nu}\ln\frac{(D-2+\nu)\lambda}{\ell}.
\end{align}
Introducing the coordinate transformation
\begin{align}
 \frac{(D-2+\nu)\lambda}{\ell}=\brac{\frac{z}{\ell}}^{D-2+\nu},
\end{align}
the solution can be reconstructed to give
\begin{align}
 \dif s^2&=-\brac{\frac{\ell}{z}}^{2\nu}\dif t^2+\frac{\ell^2}{z}\brac{\dif z^2+\dif\vec{x}^2_{(D-2)}},\quad 2\Lambda=-\frac{(D-2+\nu)(D-3+\nu)}{\ell^2},\nonumber\\
 A&=\sqrt{\frac{D-2}{2(D-3)}}\;\frac{q\ell}{D-2+\nu}\brac{\frac{z}{\ell}}^{D-2+\nu}\;\dif t,\quad q^2=\frac{(D-2+\nu)(D-3)(\nu-1)}{(D-2)\ell^2},\\
 \varphi&=-\sqrt{\frac{D-2}{D-3}}\frac{(D-2)}{\alpha}\ln\frac{z}{\ell},\quad
 \alpha^2=\frac{D-2}{\nu-1}.
\end{align}

\section{Discussion and conclusion} \label{conclusion}

By choosing a metric ansatz to have one Killing vector that is orthogonal to a product space of the form $\mathbb{R}^1\times\Sigma_{D-2}$, the Einstein-Maxwell-dilaton equations are reduced a particularly simple set of second-order ordinary differential equations. This system is equivalent to an effective Lagrangian with exponential potentials. These potentials are related to the cosmological constant $\Lambda$, Maxwell field of strength $q$, and $k$, the curvature of $\Sigma_{D-2}$.

When some of the potentials are set to zero, the equations of motion are decoupled, or can be decoupled under appropriate linear combinations of the fields. We have explored some of these cases and their associated solutions. Among the most notable solutions are perhaps the interpolating solution between the Reissner-Nordstr\"{o}m black hole and the Bertotti-Robionson spacetime, and the planar AdS naked singularity. We have also presented a one-parameter generalisation of the dilaton-Melvin solution and a rederivation of the Lifshitz spacetime which is important for holographic condensed matter physics.

The reader may have already noticed that in all the cases considered above, the solvable equations are Liouville equations of the form
\begin{align}
 \ddot{F}=K\expo{2F},
\end{align}
for some constant $K$ and $F$ being either $U$, $\Omega$, or linear combinations thereof with $\psi$. This naturally follows from the effective potentials in Eq.~\Eqref{Lagrangian} which are exponential, or Liouville potentials. In relation to this, we can foresee an immediate extension of the methods above to include specific types dilaton potentials. 

In particular, if we include Loiuville-type potentials of the form $V_0\expo{-2\kappa\varphi}$ in the action \Eqref{action}, it should still be possible to cast the resulting equations as an effective Lagrangian system with exponential potentials under the same metric and field ansatz. Other linear combinations or choices of the parameter $\kappa$ can then be chosen to find solvable systems. Indeed, there is already a rich variety of solutions to gravity with scalar fields under Liouville-type potentials \cite{Cai:1997ii,Charmousis:2001nq,Charmousis:2009xr}. In the context of our Lagrangian \Eqref{Lagrangian}, we can see that Liouville-type potentials introduce an additional parameter that can be explored to find more exact solutions.

\appendix

\section{Reduction to a one-dimensional Lagrangian} \label{reduction}

In this Appendix, we show how the ansatz \Eqref{metric_ansatz} was systematically chosen. The starting point is to maintain that the metric has at least one Killing vector $\partial_\sigma$, and therefore the metric and matter fields can be written in the form
\begin{align}
 \dif s^2&=\epsilon\expo{2U}\dif\sigma^2+\expo{-\frac{2U}{D-3}}\bar{g}_{ab}\dif y^a\dif y^b,\label{pre_metric}\\
 A&=\sqrt{\frac{D-2}{2(D-3)}}\chi\,\dif\sigma,\label{pre_A}\\
  \varphi&=\sqrt{\frac{D-2}{D-3}}\psi,\quad \alpha=\sqrt{\frac{D-3}{D-2}}a,\label{pre_phi}
\end{align}
where $U$, $\chi$, and $\psi$ depend only on $y^a$. The Einstein-Maxwell-dilaton equations become
\begin{subequations}\label{eom}
\begin{align}
 \bar{R}_{ab}=\frac{2\Lambda}{D-3}\expo{-\frac{2U}{D-3}}\bar{g}_{ab}+\frac{D-2}{D-3}\Big[\barnab_aU\barnab_b&U+\epsilon\expo{-2a\psi-2U}\barnab_a\chi\barnab_b\chi+\barnab_a\psi\barnab_b\psi\Big],\label{R_ab}\\
 \barnab^2U+\epsilon\expo{-2a\psi-2U}\brac{\barnab\chi}^2&=-\frac{2\Lambda}{D-2}\expo{-\frac{2U}{D-3}},\\
 \barnab\cdot\brac{\expo{-2a\psi-2U}\barnab\chi}&=0,\\
 \barnab^2\psi+\epsilon a\expo{-2a\psi-2U}\brac{\barnab\chi}^2&=0,
\end{align}
\end{subequations}
where $\barnab_a$ is the covariant derivative on $\bar{g}_{ab}$, with the notation $\barnab\cdot\brac{f\barnab h}=\bar{g}^{ab}\nabla_a\brac{f\nabla_b h}$, $\brac{\barnab f}^2=\bar{g}^{ab}\barnab_a f \barnab_b f$, and $\barnab^2f=\bar{g}^{ab}\barnab_a\barnab_b f$ for any functions $f=f(y^a)$ and $h=h(y^a)$. This system of equations follows from the effective action
\begin{align}
 \bar{I}&=\frac{1}{16\pi}\int\dif^{D-1}x\sqrt{|\bar{g}|}\sbrac{\bar{R}-2\Lambda\expo{-\frac{2U}{D-3}}-\frac{D-2}{D-3}\Bigl(\brac{\barnab U}^2+\epsilon\expo{-2a\psi-2U}\brac{\barnab\chi}^2+\brac{\barnab\psi}^2\Bigr)}.
\end{align}
We then specialise to the case where $\bar{g}_{ab}$ is conformal to $\mathbb{R}^1\times\Sigma_{D-2}$, with the metric
\begin{align}
 \bar{g}_{ab}\dif x^a\dif x^b=\expo{\frac{2\Omega}{D-3}}\brac{\dif z^2+\hat{h}_{ij}\dif x^i\dif x^j}, \label{barg}
\end{align}
where $\hat{h}_{ij}\dif y^i\dif y^j$ is an Einstein space of constant unit curvature $k=\pm1,0$. We assume that all metric and field functions depend only on $z$. The $(zz)$-component of Eq.~\Eqref{R_ab} is
\begin{align}
 -\frac{2(D-2)}{D-3}\frac{\dif^2\Omega}{\dif z^2}&=\frac{4\Lambda}{D-3}\expo{\frac{2\Omega-2U}{D-3}}+\frac{2(D-2)}{D-3}\brac{\brac{\frac{\dif U}{\dif z}}^2+\epsilon\expo{-2a\psi-2U}\brac{\frac{\dif\chi}{\dif z}}^2+\brac{\frac{\dif\psi}{\dif z}}^2}, \label{R_zz}
\end{align}
Taking the trace of Eq.~\Eqref{R_ab} and using Eq.~\Eqref{R_zz}, we obtain an equation of first integrals
\begin{align}
 \frac{D-2}{D-3}\Biggl[\brac{\frac{\dif U}{\dif z}}^2-\brac{\frac{\dif\Omega}{\dif z}}^2+\epsilon\expo{-2a\psi-2U}\brac{\frac{\dif\chi}{\dif z}}^2&+\brac{\frac{\dif\psi}{\dif z}}^2\Biggr]\nonumber\\
 &=2\Lambda\expo{\frac{2\Omega-2U}{D-3}}-k(D-2)(D-3). \label{constraint_z}
\end{align}
Turning to the action and integrating out the $y^i$ directions, it becomes
\begin{align}
 \bar{I}&=-\frac{(D-2)\mathrm{vol}\brac{\Sigma_{D-2}}}{16\pi(D-3)}\int\dif z\;\expo{\Omega}\bigg[\brac{\frac{\dif U}{\dif z}}^2-\brac{\frac{\dif\Omega}{\dif z}}^2+\epsilon\expo{-2a\psi-2U}\brac{\frac{\dif\chi}{\dif z}}^2\nonumber\\
   &\hspace{4cm}+\brac{\frac{\dif\psi}{\dif z}}^2-k(D-3)^2+\frac{2(D-3)}{D-2}\Lambda\expo{\frac{2\Omega-2U}{D-3}}\bigg].
\end{align}
This action can be further simplified if we remove the factor of $\expo{\Omega}$ from the effective Lagrangian. 
To this end we introduce a new coordinate via
\begin{align}
 \dif\lambda=\expo{-\Omega}\dif z,
\end{align}
and the action becomes
\begin{align}
 \bar{I}&=-\frac{(D-2)\mathrm{vol}\brac{\Sigma_{D-2}}}{16\pi(D-3)}\int\dif\lambda\bigg(\dot{U}^2-\dot{\Omega}^2+\epsilon\expo{-2a\psi-2U}\dot{\chi}^2+\dot{\psi}^2\nonumber\\
        &\hspace{6cm}-k(D-3)^2\expo{2\Omega}+\frac{2(D-3)}{D-2}\Lambda\expo{\frac{2(D-2)\Omega-2U}{D-3}}\bigg),
\end{align}
where over-dots denote derivatives with respect to $\lambda$. Solving the system now amounts to solving a dynamical system described by the Lagrangian
\begin{align}
 \mathcal{L}&=\half\sbrac{\dot{U}^2-\dot{\Omega}^2+\epsilon\expo{-2a\psi-2U}\dot{\chi}^2+\dot{\psi}^2-k(D-3)^2\expo{2\Omega}+\frac{2(D-3)}{D-2}\Lambda\expo{\frac{2(D-2)\Omega-2U}{D-3}}}.
\end{align}
Indeed, applying the Euler-Lagrange equations to $U$, $\Omega$, $\chi$, and $\psi$ gives Eq.~\Eqref{EOMS}. In terms of the coordinate $\lambda$, the constraint \Eqref{constraint_z} is
\begin{align}
 \dot{U}^2-\dot{\Omega}^2+\epsilon\expo{-2a\psi-2U}\dot{\chi}^2+\dot{\psi}^2+k(D-3)^2\expo{2\Omega}-\frac{2(D-3)}{D-2}\Lambda\expo{\frac{2(D-2)\Omega-2U}{D-3}}=0.
\end{align}
Furthermore, the metric Eq.~\Eqref{pre_metric} with $\bar{g}_{ab}$ given by Eq.~\Eqref{barg} in present coordinates is
\begin{align}
 \dif s^2&=\epsilon\expo{2U}\dif\sigma^2+\expo{\frac{2\Omega-2U}{D-3}}\brac{\expo{2\Omega}\dif\lambda^2+\hat{h}_{ij}\dif x^i\dif x^j}.
\end{align}

\bibliographystyle{emd}

\bibliography{emd}

\end{document}